\documentclass[useAMS,usenatbib]{mn2e}

\usepackage{graphicx}

\newcommand{\aap}{A\&A}
\newcommand{\apj}{ApJ}
\newcommand{\apjl}{ApJ}
\newcommand{\apjs}{ApJS}
\newcommand{\araa}{ARA\&A}
\newcommand{\mnras}{MNRAS}
\newcommand{\nat}{Nature}

\newcommand{\pasj}{PASJ}

\newcommand{\physrep}{Phys. Rep.}
\newcommand{\prd}{Phys. Rev. D}
\newcommand{\pre}{Phys. Rev. E}
\newcommand{\zap}{Z. Astrophys.}

\title[Chemothermal instability]{On the operation of the chemothermal instability in primordial star-forming clouds}

\author[Greif]{\parbox{17.5cm}{Thomas H. Greif$^{1,2}$\thanks{E-mail: tgreif@cfa.harvard.edu}, Volker Springel$^{3,4}$\vspace{0.3cm} and Volker Bromm$^5$}
\\$^1$ Harvard-Smithsonian Center for Astrophysics, 60 Garden Street, Cambridge, MA 02138, USA
\\$^1$ Max-Planck-Institut f\"{u}r Astrophysik, Karl-Schwarzschild-Stra\ss e 1, 85740 Garching, Germany
\\$^3$ Heidelberg Institute for Theoretical Studies, Schloss-Wolfsbrunnenweg 35, 69118 Heidelberg, Germany
\\$^4$ Zentrum f\"{u}r Astronomie der Universit\"{a}t Heidelberg, Astronomisches Recheninstitut, \\ M\"{o}nchhofstr. 12-14, 69120 Heidelberg, Germany
\\$^5$ Department of Astronomy and Texas Cosmology Center, University of Texas, Austin, TX 78712, USA}

\begin{document}

\maketitle
\topmargin-1cm

\begin{abstract}
We investigate the operation of the chemothermal instability in primordial star-forming clouds with a suite of three-dimensional, moving-mesh simulations. In line with previous studies, we find that the gas at the centre of high-redshift minihaloes becomes chemothermally unstable as three-body reactions convert the atomic hydrogen into a fully molecular gas. The competition between the increasing rate at which the gas cools and the increasing optical depth to H$_2$ line emission creates a characteristic dip in the cooling time over the free-fall time on a scale of $100\,{\rm au}$. As a result, the free-fall time decreases to below the sound-crossing time, and the cloud may become gravitationally unstable and fragment on a scale of a few tens of au during the initial free-fall phase. In three of the nine haloes investigated, secondary clumps condense out of the parent cloud, which will likely collapse in their own right before they are accreted by the primary clump. In the other haloes, fragmentation at such an early stage is less likely. However, given that previous simulations have shown that the infall velocity decreases substantially once the gas becomes rotationally supported, the amount of time available for perturbations to develop may be much greater than is evident from the limited period of time simulated here.
\end{abstract}


\section{Introduction}

The first stars in the Universe are believed to have formed only a few hundred million years after the big bang \citep{bl01, bl04a, cf05, glover05, glover13, loeb10}. They heated and ionized the pristine intergalactic medium \citep{bkl01, schaerer02, abs06, awb07, jgb07, yoshida07, whalen08a, greif09b}, and their supernova explosions enriched the primordial gas with the first heavy elements \citep{heger03, un03, wa08b, greif10}. They fundamentally altered the chemical and thermal state of the gas out of which the first galaxies formed, which in turn initiated the first self-sustaining cycle of star formation, feedback and chemical enrichment \citep{tw96, mf99, mfr01, oh02, rgs02a, wv03, dijkstra04, rpv06, wa08b, greif10}. Understanding the formation and properties of the first stars is thus an important step towards a comprehensive picture of structure formation in the early Universe.

The first theoretical studies of Population~III star formation date back to the late 1960's. They recognized molecular hydrogen as the sole low-temperature coolant in low-mass gas clumps that condensed out of the expanding Universe at high redshift \citep{sz67, pd68, hirasawa69, matsuda69, takeda69}. Subsequent studies used simplified one-zone models to account for the dynamics of collapsing gas clouds in the presence of radiative cooling \citep{yoneyama72, hutchins76, silk77, silk83, carlberg81, kr83, pss83, cba84, ik84, cr86, sun96, uehara96, tegmark97}, while the first one-dimensional calculations of the simultaneous collapse of dark matter (DM) and gas were carried out in the context of the cold dark matter paradigm \citep{htl96, on98, nu99}. Three-dimensional simulations of Population~III star formation had to await improvements in numerical simulation techniques, and were not possible until the late 1990`s \citep{abel98, bcl99, bcl02, abn00, abn02}.

Based on these pioneering studies, a now widely accepted `standard model' of Population~III star formation has emerged. In this picture, the first bound structures capable of hosting star formation are considered to be DM `minihaloes' with virial masses $M_{\rm vir}\ga 10^5\,{\rm M}_\odot$, which collapse at redshifts $z\ga 20$. At the centre of these haloes, molecular hydrogen forms due to the non-negligible free electron fraction left over after recombination. The ro-vibrational states of H$_2$ are excited by collisions with atomic hydrogen and helium, which decay radiatively and allow the gas to cool. The minimum temperature of about $200\,{\rm K}$ is reached at a density of $\simeq 10^4\,{\rm cm}^{-3}$, where the ro-vibrational states of H$_2$ are populated according to local thermal equilibrium \citep[LTE;][]{abel98, abn02, bcl99, bcl02}. Shortly thereafter, the central gas cloud becomes Jeans-unstable and collapses in its own right.

At densities $n_{\rm H}\ga 10^8\,{\rm cm}^{-3}$, where $n_{\rm H}$ is the number density of hydrogen nuclei, three-body reactions convert the mostly atomic gas into molecular hydrogen \citep{pss83, abn02, bl04b, yoshida06b, tao09}. During this phase, the gas may become chemothermally unstable, since the rapidly increasing H$_2$ fraction results in increased cooling, which further accelerates the collapse and the rate at which H$_2$ is formed \citep{sy77, silk83, oy03}. This runaway process is counteracted by the release of the binding energy of the formed H$_2$, and the increasing optical depth of the gas to H$_2$ line emission. As the molecular hydrogen fraction saturates, collision-induced emission becomes important, followed by collisional dissociation of H$_2$ \citep{on98, ra04}. The final, nearly adiabatic contraction of the cloud ends with the formation of a protostar with a mass of $\simeq 0.01\,{\rm M}_\odot$ \citep{on98, yoh08}.

An inherent problem of self-gravitating collapse simulations is the continually decreasing time-scale on which the gas evolves, which limits the integration time to of order a few dynamical times of the Jeans-unstable cloud. This problem may be circumvented by employing so-called sink particles, but comes at the cost of introducing artificial boundary conditions \citep{bbp95, kmk04, jappsen05, federrath10}. Recent studies that employed sink particles found that the gas becomes rotationally supported in a Keplerian disc. The disc becomes gravitationally unstable due to the high mass accretion rate from the envelope on to the disc, and the efficient cooling of the disc due to H$_2$ line emission. As a result, the disc typically fragments into a handful of protostars (Clark, Glover \& Klessen 2008; Stacy, Greif \& Bromm 2010; Clark et al. 2011a, b). A follow-up study that investigated a sample of minihaloes found that the star-forming clouds in them fragmented in a similar fashion \citep{greif11a}. An alternative approach has been taken by \citet{machida08}, where a polytropic equation of state was employed in two-dimensional simulations that used idealized initial conditions. In this case, the nature of the fragmentation depended primarily on the initial thermal and rotational energy of the cloud.

Simulations that avoid sink particles and instead directly resolve the collapse of the gas to protostellar densities have become possible, but their high computational cost limits the period of time that may be captured to $\sim 10\,{\rm yr}$  \citep{greif12}. In analogy to previous simulations with sink particles, this study finds that the circumstellar disc fragments into a small system of protostars. Strong torques exerted by the protostars and the disc result in the migration of about half of the protostars to the centre of the cloud in a free-fall time, where they merge with the primary protostar. A smaller fraction of the protostars migrates to higher orbits and survives until the end of the simulation. Two-dimensional simulations with a piece-wise polytropic equation of state show similar patterns of migration and merging \citep{vdb13}.

The susceptibility of the gas to fragmentation is a key ingredient towards understanding the initial mass function of the first stars. An early termination of fragmentation may result in the formation of a single, massive Population~III star, while continued fragmentation may lead to the formation of a rich cluster of less massive stars. Most of the mass that is accreted before the gas is dispersed by ionizing radiation resides between $10$ and $1000\,{\rm au}$ \citep{tm04, mt08, hosokawa11, sgb12}, implying that it is crucial to understand the thermal evolution of the gas on these scales. In the present study, we use a suite of three-dimensional, moving-mesh simulations to investigate the collapse of the gas on these scales in detail. Compared to previous work, we employ higher resolution and investigate a larger sample of haloes. This allows us to assess the properties of primordial gas clouds with unprecedented accuracy. In addition, we perform one of the most comprehensive resolution studies of primordial star formation to date, using up to three hundred million resolution elements.

The structure of our work is as follows. In Section~2, we describe the setup of the simulations and the chemistry and cooling network employed. In Section~3, we present a one-zone calculation that allows a detailed investigation of the chemothermal instability. We then discuss the chemothermal and gravitational stability of the gas in a representative halo in the three-dimensional simulations, followed by an analysis of the star-forming clouds in all haloes. Finally, we present a resolution study, and in Section~4 summarize our main findings and draw conclusions. All distances are quoted in proper units, unless noted otherwise.

\section{Simulations}

The three-dimensional, cosmological hydrodynamics equations for a mixed fluid consisting of collisionless DM and collisional gas are solved with the moving-mesh code {\sc arepo} \citep{springel10a}. In the following, we describe the initialization and setup of the simulations.

\subsection{Dark matter simulations}

The background Universe used to initialize the simulations is a standard $\Lambda$ cold dark matter cosmology with parameters based on the {\it Wilkinson Microwave Anisotropy Probe} observations \citep[e.g.][]{komatsu09}. We use a matter density of $\Omega_{\rm m}=1-\Omega_\Lambda=0.27$, baryon density $\Omega_{\rm b}=0.046$, Hubble parameter $h=0.7$, spectral index $n_{\rm s}=0.96$, and normalization $\sigma_8=0.81$. The matter power spectrum is evolved forward in time until $z=99$, and we use the Zel'dovich approximation to determine the initial displacements of the particles that represent the DM distribution function from a cubical lattice. We initialize nine boxes with a side length of $1\,{\rm Mpc}$ (comoving) and differing realizations of the power spectrum to investigate how our results are affected by cosmic variance. We employ $512^3$ particles of mass $\simeq 272\,{\rm M}_\odot$, and use a gravitational softening length of $\simeq 98\,{\rm pc}$ (comoving), which corresponds to $5\%$ of the initial mean interparticle separation. The simulations are evolved until the first halo grows to a virial mass of $M_{\rm vir}=5\times 10^5\,{\rm M}_\odot$, which is evaluated by an on-the-fly friends-of-friends algorithm \citep{springel01}. Various properties of these haloes are shown in Table~1.

\subsection{Resimulations}

Once the halo designated for further refinement has been located, the simulations are centred on this halo and reinitialized with higher resolution. For this purpose, the particles in the target halo and a sufficiently large boundary region around it are traced back to their initial positions, which yields the Lagrangian volume out of which the halo formed. In this region, the resolution is increased by a factor of $16^3$ with respect to the original simulation, and augmented with additional small-scale power. Each DM particle is replaced by a less massive DM particle and gas cell with a displacement corresponding to half of the initial mean interparticle separation, and a mass ratio of $M_{\rm gas}/M_{\rm dm}=\Omega_{\rm b}/(\Omega_{\rm m}-\Omega_{\rm b})$. Outside of the target region, the resolution is gradually decreased to reduce the total number of resolution elements, while the accuracy of the gravitational tidal field that influences the formation of the halo is preserved. The DM particle and gas cell masses in the high-resolution region are $\simeq 0.05$ and $\simeq 0.01\,{\rm M}_\odot$, respectively, corresponding to nearly $100$ times higher resolution compared to \citet{greif11a,greif12}. The gravitational softening length is set to $\simeq 6\,{\rm pc}$ (comoving).

The cosmological resimulations are run until the minimum cell size is of the order of $10^{-8}$ of the side length of the box. In the present version of {\sc arepo}, the number of exact integer operations used to construct a valid Voronoi tesselation increases substantially once the dynamic range exceeds the above value, which decreases the performance of the code. Nearly simultaneously, the maximum resolution of the Peano--Hilbert curve is reached and the cells can no longer be distributed efficiently among tasks. Both of these limitations will be addressed in a future version of the code. Here, we resort to extracting the central $1\,{\rm kpc}$ (comoving) of the box and reinitializing the simulations at the final output time. Once the density exceeds $n_{\rm H}=10^9\,{\rm cm}^{-3}$, we perform a second extraction and cut out the central $1\,{\rm pc}$, discard the DM particles, and evolve the simulations to a density of $n_{\rm H}=10^{15}\,{\rm cm}^{-3}$. The intermediate resimulations are necessary to achieve a maximum dynamic range in the final resimulations, which simplifies the analysis of the data. In both cases, the sound crossing time through the box is much longer than the free-fall time of the dense gas at the centre, such that perturbations from the edge of the box do not affect the central cloud.

In terms of computational efficiency, {\sc arepo} has been further improved with respect to the version employed in \citet{greif12}. One of the most important new features is a domain decomposition optimized for simulations with a high dynamic range, which allows close to optimal work balancing by taking into account the wall clock time spent on the various levels in the time step hierarchy. Furthermore, the gravitational oct-tree is now reconstructed at each time step, which prevents a dramatic increase in the wall clock time spent on the tree walk when the sizes of the nodes are adjusted dynamically to encompass all particles that are assigned to them. Finally, a separate neighbour tree is constructed to speed up the neighbour search used in particular for the construction of the Voronoi tesselation.

\begin{table}\
\centering
\caption{Redshift of collapse, virial mass, virial radius and spin parameter of the first minihaloes that form in the simulations.}
\begin{tabular}{ccccc} \hline \hline
Simulation & $z_{\rm coll}$ & $M_{\rm vir}$ & $r_{\rm vir}$ & $\lambda$ \\
& & $\left(10^5\,{\rm M}_\odot\right)$ & $\left({\rm pc}\right)$ & \\ \hline
MH1 & $23.9$ & $6.2$ & $109$ & $0.0094$ \\
MH2 & $19.2$ & $14.0$ & $176$ & $0.056$ \\
MH3 & $20.1$ & $15.5$ & $175$ & $0.019$ \\
MH4 & $24.1$ & $4.4$ & $97$ & $0.031$ \\
MH5 & $22.4$ & $5.1$ & $109$ & $0.0086$ \\
MH6 & $24.0$ & $3.5$ & $90$ & $0.053$ \\
MH7 & $24.5$ & $5.5$ & $103$ & $0.028$ \\
MH8 & $19.2$ & $6.3$ & $135$ & $0.050$ \\
MH9 & $25.0$ & $3.1$ & $83$ & $0.031$ \\ \hline
\multicolumn{5}{l}{} \\
\end{tabular}
\end{table}

\subsection{Refinement}

The so-called Truelove criterion states that at least four cells per Jeans length are necessary to resolve gravitational instability \citep{truelove98}. However, this is generally not sufficient to resolve the turbulent cascade that is driven by gravitational collapse. For example, \citet{federrath11} showed that at least $32$ cells per Jeans length are necessary to model the amplification of magnetic fields by the turbulent dynamo. In \citet{turk12}, it was shown that the employed resolution affects the chemical, thermal and dynamical evolution of primordial gas clouds. In particular, the amount of rotational support increases as the resolution is decreased, which might affect the ability of the gas to fragment. Although the resolution requirements in a fixed grid simulation may not be the same as in a moving-mesh simulation, these results show that it is desirable to use as many cells per Jeans length as possible.

\citet{tna10} further found that it is advantageous to evaluate the Jeans length using the minimum temperature of the gas, which ensures that shock-heated regions are equally well resolved as cold regions. To facilitate a comparison with previous studies, we therefore also employ the refinement criterion proposed by \citet{tna10}, and evaluate the Jeans length at $T_{\rm min}=200\,{\rm K}$, which yields:
\begin{equation}
\lambda_{\rm J}\simeq0.6\,{\rm pc}\left(\frac{n_{\rm H}}{10^4\,{\rm cm}^{-3}}\right)^{-1/2}.
\end{equation}
Using this definition, we refine cells if $h>\lambda_{\rm J}/N_{\rm J}$, where $h=(3V/4\pi)^{1/3}$ is the approximate radius of a cell, $V$ its volume, and $N_{\rm J}$ the desired number of cells per Jeans length. For $N_{\rm J}=32$ and $n_{\rm H}\ga 10^8\,{\rm cm}^{-3}$, the Jeans mass is resolved by $\simeq 10$ times more cells compared to \citet{greif12}. In Section~3.3, we present a resolution study in which we systematically increase $N_{\rm J}$ from $8$ to $64$. For the main simulations discussed in Sections~3.1 and 3.2, we use $64$ cells per Jeans length.

In addition to the refinement based on the Jeans length, we refine cells if their mass increases to more than twice their initial mass. While {\sc arepo} maintains near constant mass resolution by advecting the cells in a Lagrangian fashion, the Eulerian mass flux between cells may lead to a gradual drift away from the initial cell masses.

\begin{figure}
\begin{center}
\includegraphics[width=8cm]{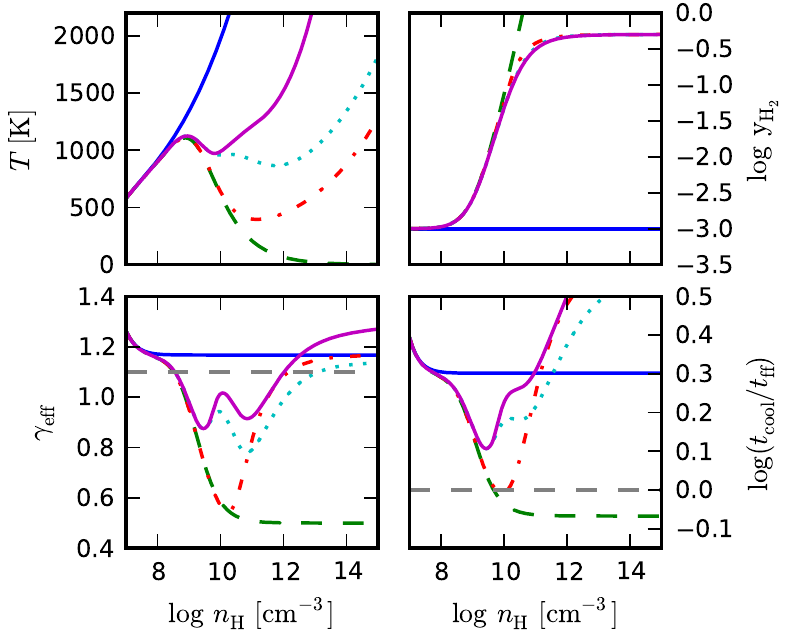}
\caption{From top left to bottom right: temperature, H$_2$ fraction, effective equation of state, and cooling time over free-fall time versus number density of hydrogen nuclei in a one-zone calculation. The various profiles show the evolution for a constant H$_2$ fraction (blue), indefinitely increasing H$_2$ fraction (green), limited H$_2$ fraction (red), three-body formation heating (cyan) and an increasing optical depth to line emission (magenta). The dashed grey lines denote an effective equation of state of $1.1$ and a cooling time equal to the free-fall time. If the H$_2$ fraction is allowed to increase indefinitely, the chemothermal instability results in an asymptotic $\gamma_{\rm eff}=1/2$. However, the decreasing H$_2$ fraction, three-body formation heating, and the optical depth of the gas reduce the effectiveness of the instability, such that the cooling time over the free-fall time drops only slightly below the value expected for a constant H$_2$ fraction.}
\label{fig_onezone}
\end{center}
\end{figure}

\subsection{Chemistry and cooling}

The chemistry and cooling network employed in the present study is nearly identical to the one described in \citet{greif12}. The species evolved next to the internal energy of the gas are H, H$^+$, H$^-$, H$_2^+$, H$_2$, He, He$^+$, He$^{++}$, D, D$^+$, HD, and free electrons \citep{gj07, clark11a}. We use the publicly available {\sc sundials cvode} package to solve the coupled chemical and thermal rate equations (Hindmarsh et al. 2005). The abundances are computed with a relative accuracy of $0.1\%$, which is necessary for an accurate integration at high densities, where three-body reactions become important. The absolute tolerance is set to $y_X=10^{-20}$, where $y_X$ denotes the number density of species $X$ with respect to the number density of hydrogen nuclei. This threshold prevents the solver from computing accurate abundances when they have become low enough that they may be considered unimportant. In contrast to \citet{greif12}, we do not switch to an equilibrium chemistry solver at a density of $n_{\rm H}=10^{14}\,{\rm cm}^{-3}$, since we only follow the collapse up to a density of $n_{\rm H}=10^{15}\,{\rm cm}^{-3}$.

The most important low-temperature coolant in primordial gas is molecular hydrogen, which forms via associative detachment of H and H$^-$, following radiative association of H and e$^-$ \citep{pd68, hat69, abel97, gp98}. The ro-vibrational states of H$_2$ are collisionally excited and decay radiatively, allowing the gas to cool to a minimum temperature of $T\simeq 200\,{\rm K}$. At low densities, we use the excitation rates for collisions with neutral hydrogen atoms of \citet{wf07}, assuming an ortho:para ratio of 3:1. At high densities, when the energy states are populated according to LTE, we use the rates of \citet{lpf99}. We neglect collisions with electrons, protons, and neutral helium atoms, which may provide a non-negligible amount of cooling at the relevant densities and temperatures \citep{ga08}.

Another low-temperature coolant in primordial gas is hydrogen deuteride (HD), which may become important at temperatures $T\la 200\,{\rm K}$ due to its permanent dipole moment \citep{slp98, flower00, ui00, nu02, no05, jb06, ripamonti07, yoh07, mb08, greif11a}. In minihaloes, HD forms mainly via associative detachment of H$_2$ and D$^+$, where the abundance of D$^+$ is set predominantly by recombinations and charge transfer with H. We follow \citet{lna05} for the cooling rate of the gas by HD ro-vibrational transitions.

\begin{figure}
\begin{center}
\includegraphics[width=8cm]{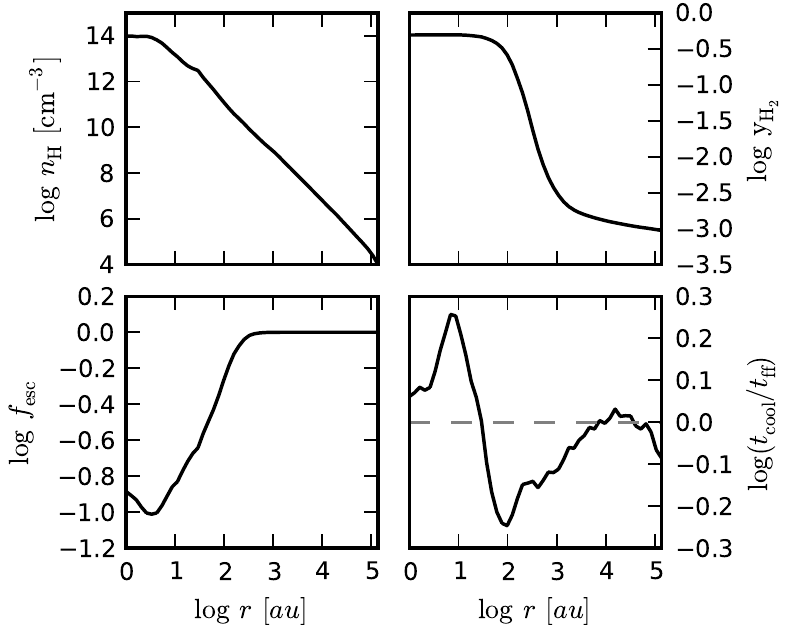}
\caption{From top left to bottom right: number density of hydrogen nuclei, H$_2$ fraction, photon escape fraction, and cooling time over free-fall time versus radius. The dashed grey line denotes a cooling time equal to the free-fall time. The H$_2$ fraction rapidly increases below $\simeq 1000\,{\rm au}$ due to three-body reactions, which results in the cooling time falling below the free-fall time. As the optical depth of the gas to H$_2$ line emission increases, the cooling time once again becomes larger than the free-fall time.}
\label{fig_radial_a}
\end{center}
\end{figure}

At densities $n_{\rm H}\ga 10^8\,{\rm cm}^{-3}$, molecular hydrogen forms via three-body reactions, which heats the gas due to the release of the molecular binding energy \citep{pss83, bcl99, abn02,bcl02,yoshida06b}. The corresponding reaction rates are highly uncertain, which is reflected by the substantial variation of the properties of primordial gas clouds for different rates \citep{glover08, turk11}. We here adopt the intermediate rate for three-body H$_2$ formation among those of \citet{turk11}, which is taken from \citet{pss83}. For $n_{\rm H}\ga 10^{10}\,{\rm cm}^{-3}$, column densities are high enough that the gas becomes optically thick to H$_2$ line emission, and we use the Sobolev approximation to determine the escape fraction of photons emitted by H$_2$ lines \citep{yoshida06b, clark11a}.

\subsection{Analysis}

The number of resolution elements employed in the main simulations is $\simeq 3\times 10^8$, which is an increase by up to two orders of magnitude compared to the simulations presented in \citet{greif11a,greif12}. Such large data sets also require more sophisticated tools for an efficient analysis. To this end, we have written the parallel software package {\sc sator}, which uses the message passing interface (MPI) to read data sets in {\sc c} and distribute the particle load to the processing cores. The operations required for the various tasks, e.g. the image generation or the computation of radial profiles, are then performed on the particles residing on each core, allowing near-optimal work balance. Upon completion, an intermediate data set is written that is pipelined to a {\sc python} program for further analysis. The combined workflow of {\sc sator} is controlled by a script that reads in a minimal set of parameters from the command line, which specifies the mode of operation and the target data set. The script may be run interactively, or submitted to a queuing system for very large data sets and time-consuming calculations. On a Sandy bridge computing cluster, the runtime of {\sc sator} to analyse a data set with $\simeq 3\times 10^8$ cells using $64$ MPI tasks is of the order of $10\,{\rm s}$, which is a substantial improvement compared to the performance of the analysis tools used in \citet{greif11a, greif12}.

\begin{figure}
\begin{center}
\includegraphics[width=8cm]{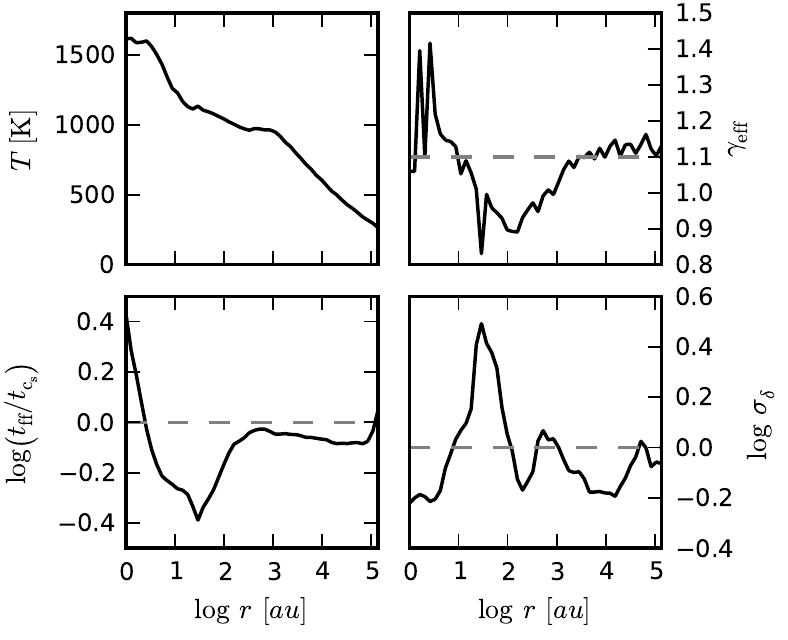}
\caption{From top left to bottom right: temperature, effective equation of state, free-fall time over sound-crossing time and rms density contrast versus radius. The dashed grey lines denote an effective equation of state of $1.1$, a free-fall time equal to the sound-crossing time and an rms density contrast of unity. As the cooling time falls below the free-fall time, the effective equation of state of the gas drops from $\gamma_{\rm eff}\simeq 1.1$ to $\simeq 0.9$, and the temperature remains nearly constant at about $1000\,{\rm K}$ between $\simeq 1000$ and $\simeq 10\,{\rm au}$. On a scale of a few tens of au, the free-fall time falls below the sound-crossing time, and the central gas cloud becomes gravitationally unstable. The resulting growth of overdensities is evident from the bottom-right panel, which shows that the rms density contrast increases to well above unity.}
\label{fig_radial_b}
\end{center}
\end{figure}

\begin{figure*}
\begin{center}
\includegraphics[width=16cm]{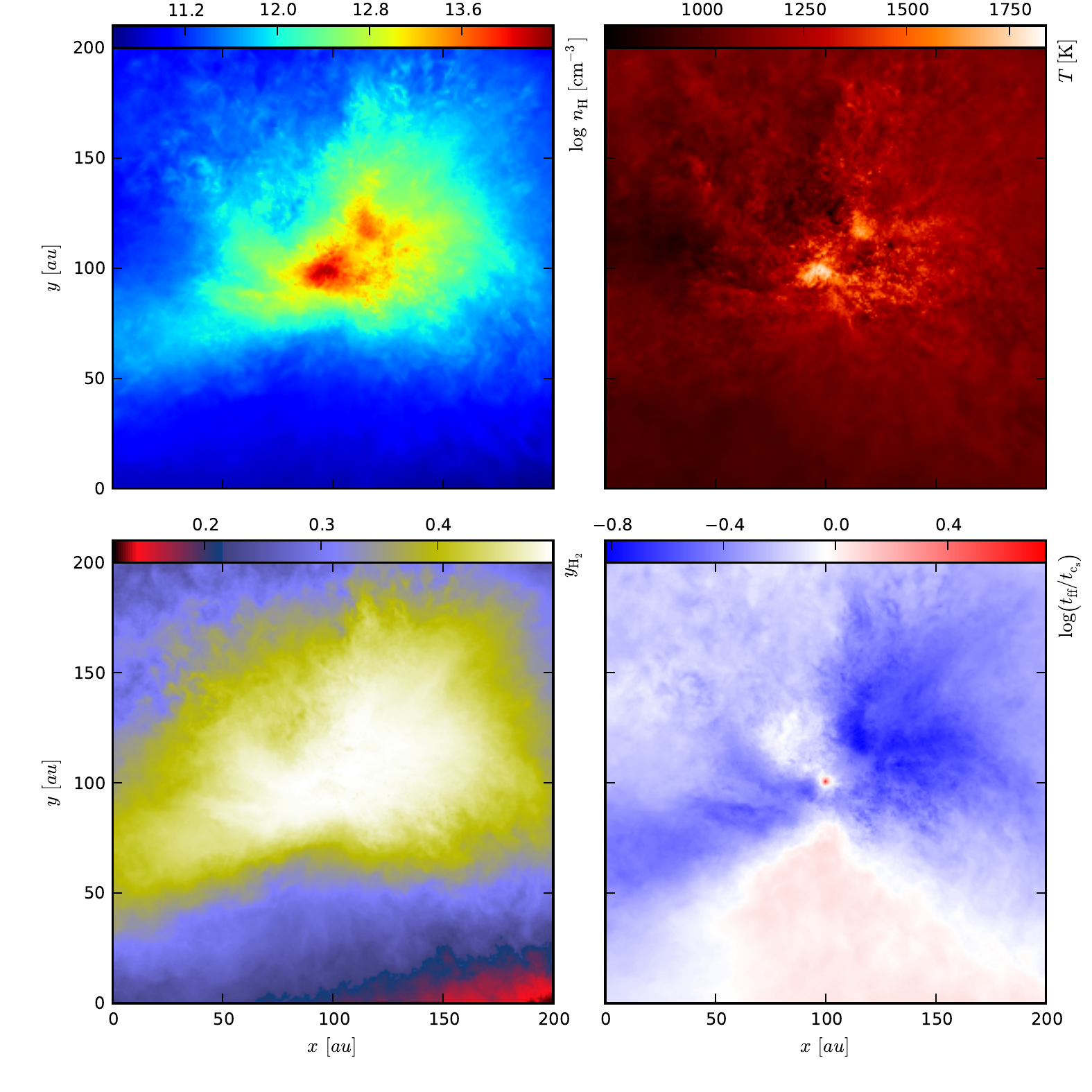}
\caption{From top left to bottom right: number density of hydrogen nuclei, temperature, H$_2$ fraction and free-fall time over sound-crossing time in a cube of side length $200\,{\rm au}$, weighted with the mass and the square of the density of the cells traversed along the line of sight. As the chemothermal instability develops and the cooling time falls below the free-fall time, a secondary clump condenses out of the parent cloud, which is visible to the top right of the primary clump. The gravitational instability of the clump is evident from the bottom-right panel, which shows that the free-fall time has dropped below the sound-crossing time.}
\label{fig_img}
\end{center}
\end{figure*}

\section{Results}

\subsection{Chemothermal and gravitational instability}

The classical chemothermal instability as envisaged by \citet{yoneyama73} is an extension of the \citet{field65} thermal instability to gases with a variable chemical composition. Based on this model, \citet{sy77} investigated the growth of perturbations in a primordial, spherically symmetric gas cloud contracting on the free-fall time-scale, in which the H$_2$ fraction increases due to three-body reactions. They found that a range of densities and temperatures exists where perturbations grow, which has been confirmed by subsequent one- and three-dimensional studies \citep{ra04, yoshida06b}. Crucial assumptions in deriving the governing equations are that the gas behaves isobarically, which excludes the influence of pressure gradients, gravity, and turbulence. These caveats prompted us to investigate the operation of the chemothermal instability and the resulting gravitational instability from a different perspective. Instead of a `bottom-up' process stemming from the growth of infinitesimal perturbations, we suggest that the potential fragmentation that occurs in primordial gas clouds is explained by a `top-down' gravitational instability, which is triggered by a chemothermal instability.

To understand the origin and development of these instabilities in detail, we will first use a one-zone model to demonstrate that the rapidly increasing H$_2$ fraction results in the cooling time falling below the free-fall time. We then extend our analysis to the full three-dimensional simulations, and show that the chemothermal instability may trigger gravitational instability and fragmentation.

\subsubsection{One-zone model}

In a spherically symmetric, roughly isothermally contracting cloud (an appropriate assumption in the regime at which the chemothermal instability operates), the rate at which the core density increases is given by
\begin{equation}
{\dot \rho}=\rho/t_{\rm ff},
\end{equation}
where $\rho=m_{\rm H}n_{\rm H}/X$ is the mass density, $m_{\rm H}$ is the mass of the hydrogen atom, $X=3/4$ is the cosmological mass fraction of hydrogen, $t_{\rm ff}=\sqrt{3\pi /\left(32G\rho\right)}$, and $G$ is Newton's constant. The volumetric heating rate of the gas in ${\rm erg}\,{\rm s}^{-1}\,{\rm cm}^{-3}$ is given by
\begin{equation}
{\dot u}=\gamma u/t_{\rm ff}-Ay_{\rm H_2}f_{\rm esc}\left(\frac{n_{\rm H}}{{\rm cm}^{-3}}\right)\left(\frac{T}{{\rm K}}\right)^\alpha +Bn_{\rm H}{\dot y_{\rm H_2}},
\end{equation}
where $\gamma$ is the adiabatic index of the gas, $y_{\rm H_2}$ the molecular hydrogen fraction, $A=10^{-33}\,{\rm erg}\,{\rm cm}^{-3}\,{\rm s}^{-1}$ and $\alpha=4$ approximate the H$_2$ line cooling function used in the full three-dimensional simulations, $B=4.48\,{\rm eV}$ is the binding energy of the H$_2$ molecule and $f_{\rm esc}$ is the photon escape fraction for H$_2$ line emission. The first term on the right-hand side of the equation describes adiabatic heating, the second term H$_2$ line emission, and the third term three-body formation heating. We have verified that all other thermal processes are comparatively subdominant at the relevant densities and temperatures. We use an adiabatic index of $\gamma=5/3$ for a monatomic gas, even though the gas consists of a mix of atomic and molecular hydrogen. We also neglect the change in the mean molecular weight $\mu$, which counteracts the effect of the varying $\gamma$ in the relation between the energy density and temperature. The net effect of neglecting both processes is therefore very small. The photon escape fraction is given by
\begin{equation}
f_{\rm esc}=\frac{ax}{x^{a}+a-1}
\end{equation}
for $x\ge 1$ and $f_{\rm esc}=1$ for $x<1$, where $x=n_{\rm H}/n_{\rm H, thresh}$, $n_{\rm H, thresh}=4\times 10^9\,{\rm cm}^{-3}$ and $a=1.45$. This formula is based on the fit provided by \citet{ra04}, but has the advantage of a continuous derivative at $x=1$. Finally, the rate at which H$_2$ molecules are formed is given by
\begin{equation}
{\dot y_{\rm H_2}}=C\left(1-2y_{\rm H_2}\right)^3\left(\frac{n_{\rm H}}{{\rm cm}^{-3}}\right)^2\left(\frac{T}{{\rm K}}\right)^{-1},
\end{equation}
where $C=5.5\times 10^{-29}\,{\rm s}^{-1}$. We have again verified that all other rate equations that affect the chemical composition of the gas are comparatively small at the relevant densities and temperatures. We initialize the calculations at a density of $n_{\rm H}=10^7\,{\rm cm}^{-3}$, using a temperature of $T=580\,{\rm K}$ and an H$_2$ fraction of $y_{\rm H_2}=10^{-3}$. These initial conditions roughly correspond to the expected values at the initial density, and in addition yield a smooth transition into the regime at which the chemothermal instability operates.

In Fig.~\ref{fig_onezone}, we show the results of the one-zone calculation as we systematically increase the level of sophistication. The blue line shows the evolution for a constant H$_2$ fraction and a constant escape fraction of unity. In this case, the effective equation of state remains close to $\gamma_{\rm eff}=7/6$, since $t_{\rm cool}\simeq t_{\rm ff}$, which yields $T^{-3}\propto n_{\rm H}^{-1/2}$ and thus $T\propto\rho^{1/6}$. The green line shows the collapse assuming that the H$_2$ fraction may increase indefinitely, the three-body formation rate is independent of the H$_2$ fraction, and that three-body formation heating as well as the influence of the decreasing escape fraction may be neglected. In this case, an asymptotic $\gamma_{\rm eff}=1/2$ is reached, which is evident from a simple calculation.

The cooling time due to H$_2$ line emission in the optically thin limit is given by
\begin{equation}
t_{\rm cool}\propto y_{\rm H_2}^{-1}T^{-3},
\end{equation}
and the derivative of the H$_2$ fraction with respect to $n_{\rm H}$ is given by
\begin{equation}
\frac{{\rm d}y_{\rm H_2}}{{\rm d}n_{\rm H}}=\frac{{\rm d}y_{\rm H_2}}{{\rm d}t}\frac{{\rm d}t}{{\rm d}n_{\rm H}}\propto n_{\rm H_2}^{3/2-\gamma_{\rm eff}},
\end{equation}
where we have assumed $T\propto\rho^{\gamma_{\rm eff}-1}$. Integration over $n_{\rm H_2}$ and insertion into equation~(6) yields:
\begin{equation}
t_{\rm cool}\propto n_{\rm H_2}^{3/2-2\gamma_{\rm eff}},
\end{equation}
and the cooling time over the free-fall time is thus given by:
\begin{equation}
\frac{t_{\rm cool}}{t_{\rm ff}}\propto n_{\rm H_2}^{1-2\gamma_{\rm eff}}.
\end{equation}
A constant ratio of the cooling time to the free-fall time is therefore only possible for $\gamma_{\rm eff}=1/2$, which agrees with the results shown in Fig.~\ref{fig_onezone}.

\begin{figure}
\begin{center}
\includegraphics[width=8cm]{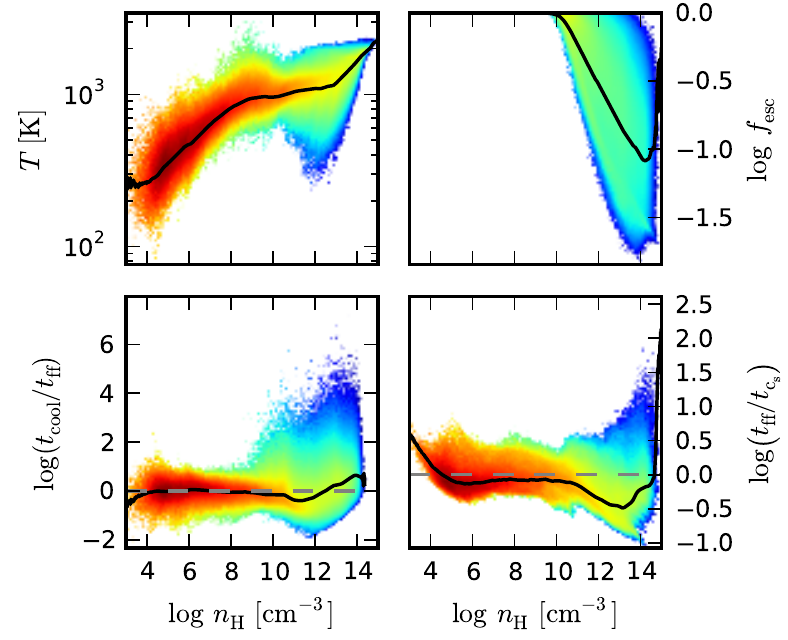}
\caption{From top left to bottom right: temperature, escape fraction, cooling time over free-fall time, and free-fall time over sound-crossing time versus number density of hydrogen nuclei. The logarithm of the mass per bin over the total mass in the box is colour coded from blue (lowest) to red (highest). The solid lines show the mass-weighted average value. The grey dashed lines denote a cooling time equal to the free-fall time, and a free-fall time equal to the sound-crossing time. The temperature distribution widens significantly above $n_{\rm H}\simeq 10^{10}\,{\rm cm}^{-3}$ as the cooling time falls below the free-fall time. The photon escape fraction at a given density fluctuates by up to two orders of magnitude, such that some regions of the cloud cool very efficiently and may become Jeans unstable. This is reflected by the substantial amounts of gas with a free-fall time below the sound-crossing time.}
\label{fig_pspace}
\end{center}
\end{figure}

Under more realistic circumstances, this asymptotic value is never achieved, since only a finite amount of H$_2$ may be formed. In addition, the rate at which H$_2$ forms decreases as the H$_2$ abundance increases. The red line in Fig.~\ref{fig_onezone} denotes the case in which these considerations have been taken into account. As expected, the effective equation of state departs from the asymptotic $\gamma_{\rm eff}=1/2$ as the H$_2$ abundance increases to of the order of unity, and quickly returns to $\gamma_{\rm eff}=7/6$. When three-body formation heating and the increasing optical depth to H$_2$ line emission are taken into account (cyan and magenta lines, respectively), the effectiveness of the chemothermal instability is further reduced. Nevertheless, an unstable regime develops at a density of $\ga n_{\rm H}\simeq 10^9\,{\rm cm}^{-3}$, where the cooling time drops below the free-fall time. This thermal instability may later trigger gravitational instability.

\begin{figure*}
\begin{center}
\includegraphics[width=16cm]{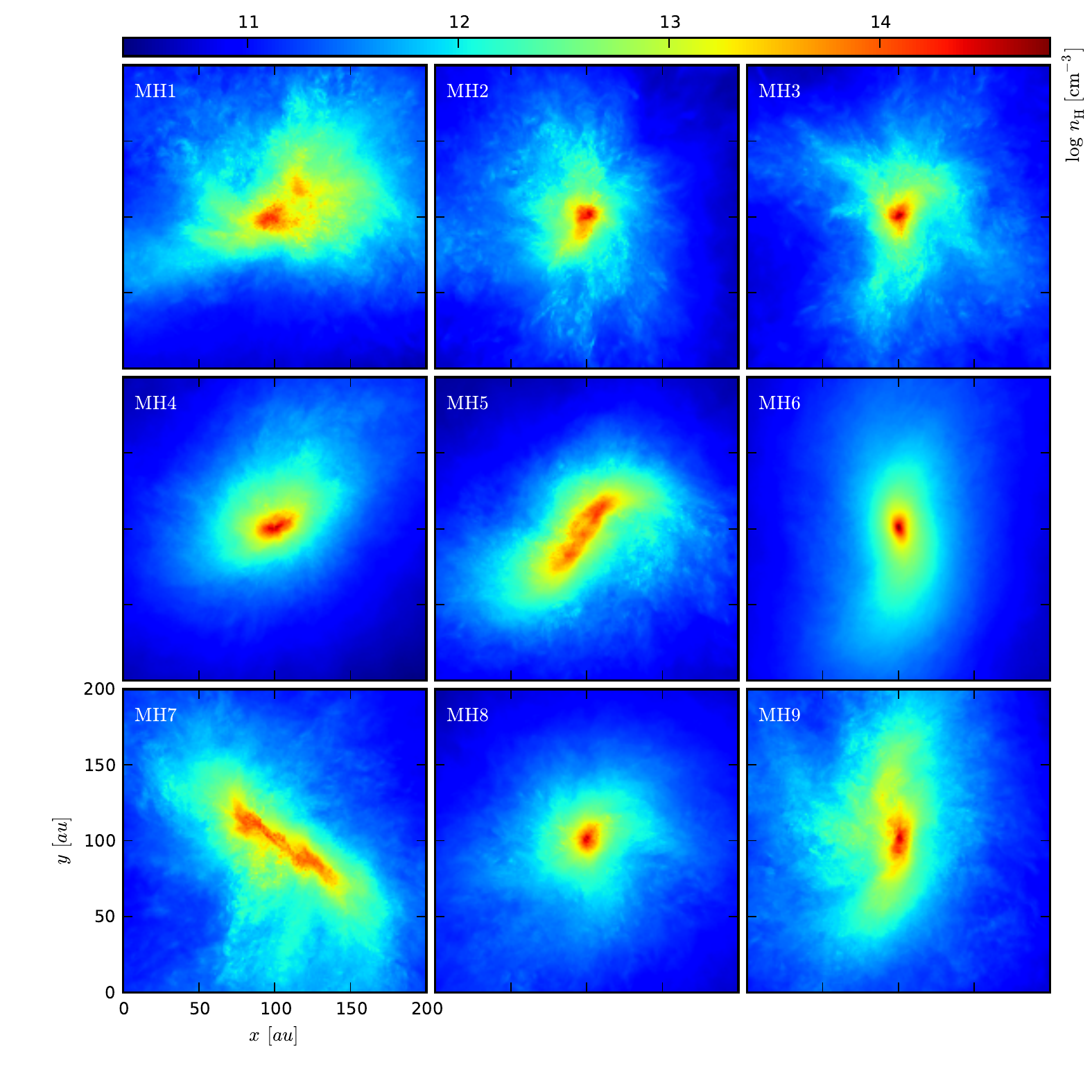}
\caption{Number density of hydrogen nuclei in cubes of side length $200\,{\rm au}$, centred on the densest cell in each halo. The number density is weighted with the mass and the square of the density of the cells traversed along the line of sight. Overall, the clouds are centrally concentrated, but display very irregular morphologies with filaments and knots indicative of turbulence. In MH5 and MH7, elongated clouds have formed, while in MH6 a prominent disc with little substructure is present. In MH1, MH5 and MH7, the collapse of a secondary clump has progressed far enough that it is clearly discernible from the first.}
\label{fig_multi_dens}
\end{center}
\end{figure*}

\subsubsection{Simulations}

The properties of the gas in the three-dimensional simulations are similar, but not entirely equal to those in the one-zone calculations. In Fig.~\ref{fig_radial_a}, we show the number density of hydrogen nuclei, H$_2$ fraction, photon escape fraction and cooling time over free-fall time versus radius $r$ at the last output time of the high-density simulations in MH1. The profiles are determined by using a mass-weighted average of the cells that contribute to the radial bins. The top-right panel shows that the H$_2$ fraction increases rapidly on a scale of $\simeq 1000\,{\rm au}$ as three-body reactions become important. In analogy to the one-zone calculations, the cooling time over the free-fall time then decreases by about a factor of $2$, and in the simulation even falls below the free-fall time. On a scale of $\simeq 100\,{\rm au}$, the optical depth of the gas to H$_2$ line emission increases, and the cooling time again approaches the free-fall time.

\begin{figure*}
\begin{center}
\includegraphics[width=16cm]{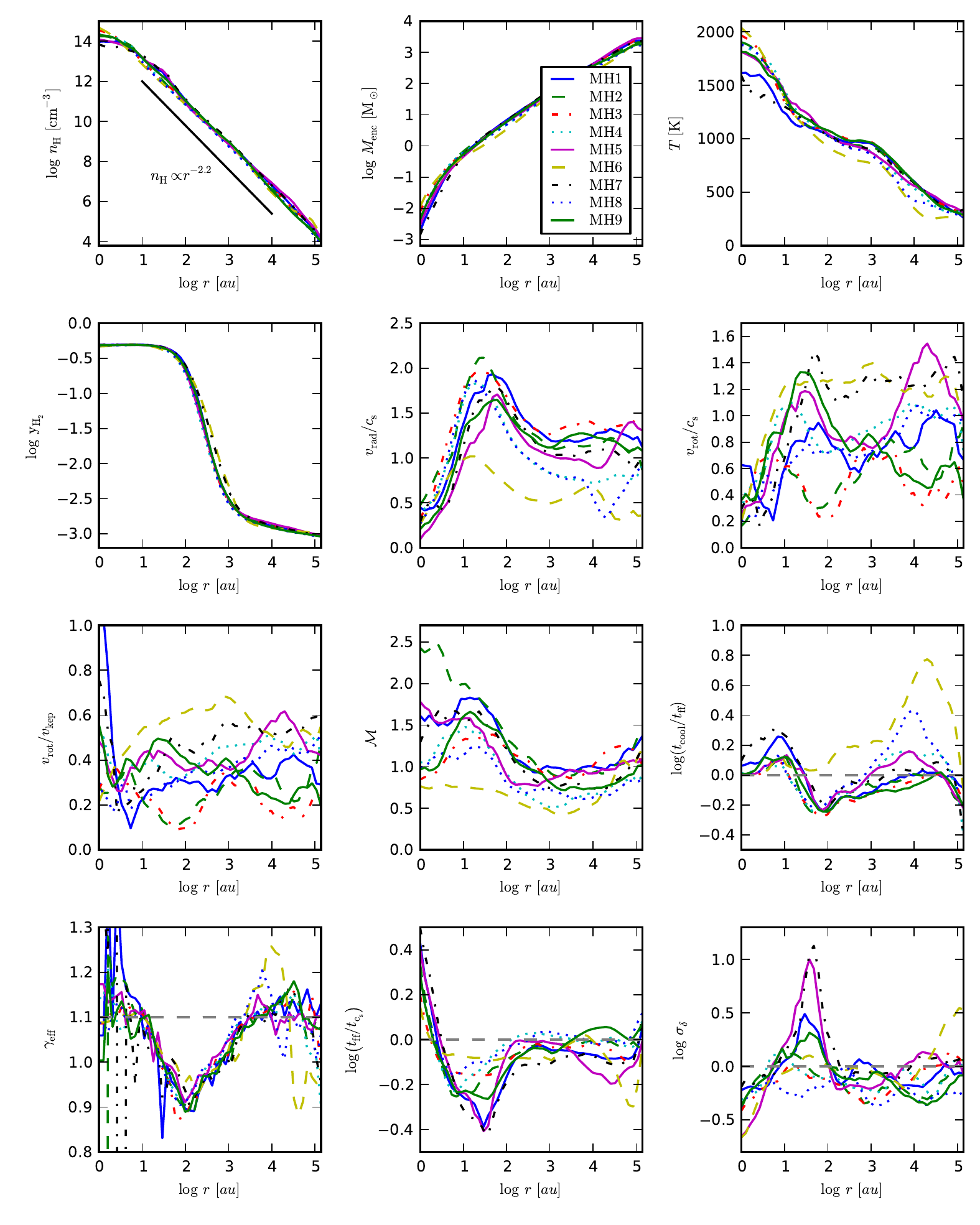}
\caption{From top left to bottom right: number density of hydrogen nuclei, enclosed gas mass, temperature, H$_2$ fraction, radial velocity over sound speed, rotation velocity over sound speed, rotation velocity over Keplerian velocity, turbulent Mach number, cooling time over free-fall time, effective equation of state, free-fall time over sound-crossing time and rms density contrast versus radius. The solid black line in the top-left panel denotes an $r^{-2.2}$ density profile. The grey dashed lines denote a cooling time equal to the free-fall time, an effective equation of state of $1.1$, a free-fall time equal to the sound-crossing time and an rms density contrast of unity. A detailed discussion of this figure is provided in Section~3.2.}
\label{fig_multi_radial}
\end{center}
\end{figure*}

\begin{figure*}
\begin{center}
\includegraphics[width=16cm]{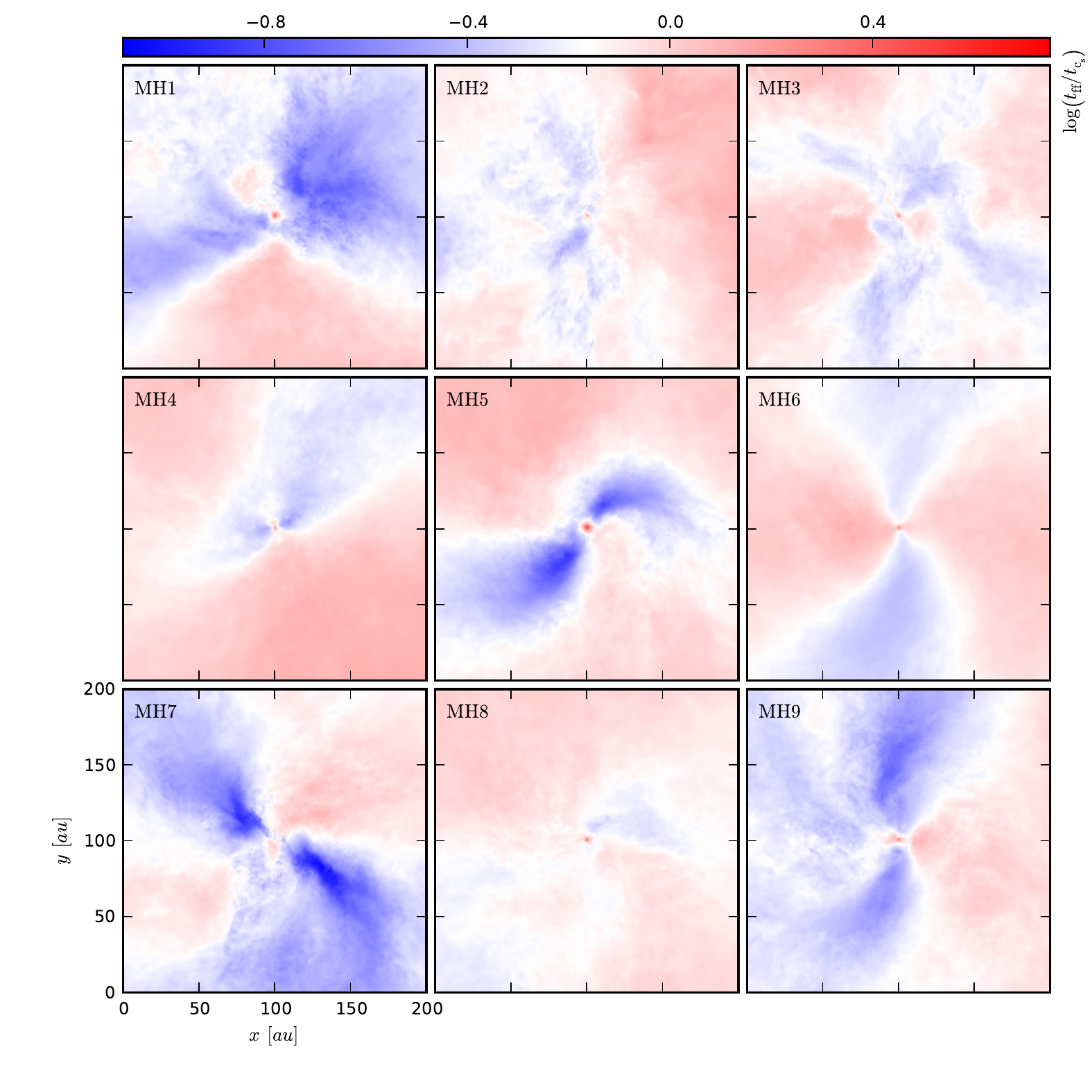}
\caption{Free-fall time over sound-crossing time in cubes of side length $200\,{\rm au}$, centred on the densest cell in each halo and weighted with the mass and the square of the density of the cells traversed along the line of sight. The most gravitationally unstable parcels of gas are visible in MH1, MH5 and MH7.}
\label{fig_multi_collapse}
\end{center}
\end{figure*}

In Fig.~\ref{fig_radial_b}, we show the temperature, effective equation of state, free-fall time over sound-crossing time and root-mean-square (rms) density contrast versus radius. The latter is given by:
\begin{equation}
\sigma_\delta=\sqrt{\sum_i \frac{m_i}{M_{\rm tot}}\left(\frac{\rho_i-{\bar\rho}}{\bar\rho}\right)^2},
\end{equation}
where the sum extends over all cells contributing to a radial bin, $i$ denotes the cell under consideration, $m_i$ its mass, $\rho_i$ its density, $M_{\rm tot}$ the total mass per bin, and ${\bar\rho}$ the mass-weighted average density of the bin. Following the decrease of the cooling time below the free-fall time, the top left panel in Fig.~\ref{fig_radial_b} shows that the temperature of the gas remains roughly constant at $\simeq 1000\,{\rm K}$ over nearly two orders of magnitude in density, which is reflected in the drop of the effective equation of state from $\gamma_{\rm eff}\simeq 1.1$ to $\simeq 0.9$. The temperature does not rise as rapidly as in the one-zone calculation, which is most likely attributable to the higher photon escape fraction obtained with the Sobolev method as opposed to a simple density dependence \citep{turk11}.

Following the thermal instability of the gas characterized by $t_{\rm cool}\la t_{\rm ff}$, the sound-crossing time does not decrease as rapidly as the free-fall time, which results in the central gas cloud becoming gravitationally unstable. The minimum in the free-fall time over the sound-crossing time develops on scales below the minimum in the cooling time over the free-fall time, since a significant change in the sound speed requires of order a cooling time, which drops to about $t_{\rm ff}/2$ (see Fig.~\ref{fig_radial_a}). This is reflected in the minimum of the free-fall time over the sound-crossing time located at a few tens of au, while the minimum in the cooling time over the free-fall time is located at $\simeq 100\,{\rm au}$. The elevated density contrast that develops during this phase is evident from the bottom-right panel, which shows that the rms density contrast increases to well above unity on a scale of a few tens of au.

The tentative formation of a secondary clump resulting from the gravitational instability of the gas is evident from Fig.~\ref{fig_img}, which shows the number density of hydrogen nuclei, temperature, H$_2$ fraction and free-fall time over sound-crossing time in the central $200\,{\rm au}$ of the box. The plane of view is perpendicular to the net angular momentum vector of the gas. The secondary clump is located to the top right of the primary clump, and its gravitational instability is apparent from the bottom-right panel. In Section~3.3, we investigate whether this clump has acquired enough mass to become Jeans unstable in its own right, and whether it collapses before it is accreted by the primary clump.

\begin{figure}
\begin{center}
\includegraphics[width=8cm]{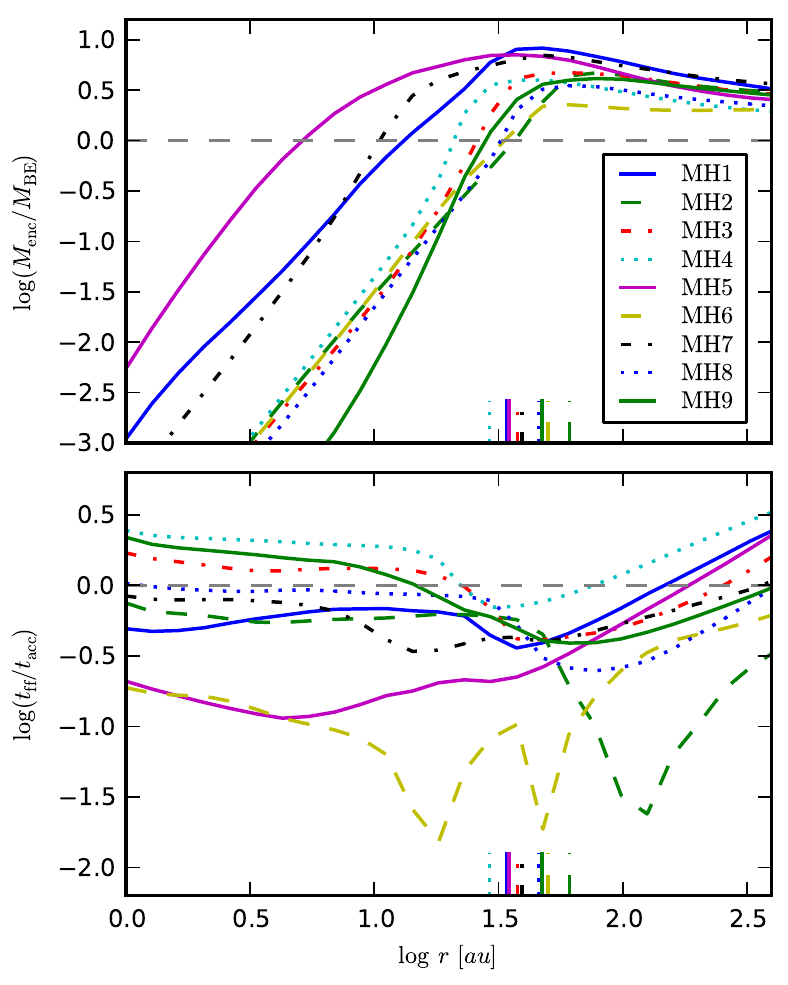}
\caption{Enclosed mass over Bonnor--Ebert mass (top panel), and free-fall time over accretion time (bottom panel) versus radius, centred on the gravitationally most unstable regions found in Fig.~\ref{fig_multi_collapse}. The vertical lines denote the separation of the clumps from the centre of the cloud. The grey dashed lines denote an enclosed mass equal to the Bonnor--Ebert mass and a free-fall time equal to the accretion time. MH1, MH5 and MH7 are likely to fragment during the initial free-fall phase, since the radius at which the enclosed mass exceeds the Bonnor--Ebert mass and the free-fall time falls below the sound-crossing is significantly smaller than the separation of the clumps from the centre of the cloud. This is not the case in the other haloes, where the primary and secondary clumps are nearly indistinguishable.}
\label{fig_mbe}
\end{center}
\end{figure}

In Fig.~\ref{fig_pspace}, we show the temperature, escape fraction, cooling time over free-fall time and free-fall time over sound-crossing time versus number density of hydrogen nuclei. The logarithm of the mass per bin over the total mass in the box is colour coded from blue (lowest) to red (highest), and the solid lines show the mass-weighted average value versus density. As the chemothermal instability develops, the temperature distribution widens and some parcels of gas cool to as low as $\simeq 300\,{\rm K}$, while others are heated to over $2000\,{\rm K}$. The photon escape fraction at a given density fluctuates by up to two orders of magnitude, showing that in the three-dimensional simulations some regions of the cloud cool very efficiently, while in the one-zone model they would lie well within the optically thick regime. These regions may become Jeans unstable, which is reflected by the abundance of gas with a free-fall time below the sound-crossing time. Another noteworthy feature is the shock-heating of a parcel of gas to over $3000\,{\rm K}$, which is apparent from the peak in temperature at $n_{\rm H}\simeq 10^9\,{\rm cm}^{-3}$. The elevated temperature results in collisional dissociation of H$_2$ down to an abundance of $y_{\rm H_2}\simeq 10^{-6}$. Such shock-heated parcels of gas are also present in the simulations of \citet{tna10}.

\subsection{Halo sample}

In Fig.~\ref{fig_multi_dens}, we show the number density of hydrogen nuclei in a face-on view of the central $200\,{\rm au}$ in the various haloes investigated. Most clouds display a centrally concentrated, irregular morphology with a multitude of filaments and knots, while MH4 and MH6 have a more regular morphology and less substructure. In MH5 and MH7, an elongated cloud has formed, while in MH6 a prominent disc with little substructure is present. In MH1, MH5 and MH7, the collapse of a secondary clump has progressed far enough that it is clearly discernible from the first.

Fig.~\ref{fig_multi_radial} shows radially averaged profiles of various physical quantities. From top left to bottom right, the panels show the number density of hydrogen nuclei, enclosed gas mass, temperature, H$_2$ fraction, radial velocity over sound speed, rotation velocity over sound speed, rotation velocity over Keplerian velocity, turbulent Mach number, cooling time over free-fall time, effective equation of state, free-fall time over sound-crossing time and rms density contrast versus radius. The rotation velocity vector ${\bmath v}_{\rm rot}$ is given by the sum of the angular momenta of the contributing cells divided by the total mass $M_{\rm tot}$ and radius of the bin. The Keplerian velocity is given by $v_{\rm kep}=\sqrt{GM_{\rm enc}/r}$, where $M_{\rm enc}$ is the enclosed mass. The turbulent velocity is given by
\begin{equation}
v_{\rm turb}=\frac{\sqrt{\sum_i m_i\left({\bmath v}_i-{\bmath v}_{\rm rad}-{\bmath v}_{\rm rot}\right)^2}}{M_{\rm tot}},
\end{equation}
where ${\bmath v}_i$ is the velocity vector of a cell, and ${\bmath v}_{\rm rad}$ the radial velocity vector of the bin. The turbulent Mach number $\mathcal{M}$ is given by $\mathcal{M}=v_{\rm turb}/c_{\rm s}$, where $c_{\rm s}$ is the sound speed.

The top-left panel in Fig.~\ref{fig_multi_radial} shows that the gas clouds have an approximate $r^{-2.2}$ profile, which is expected for $\gamma_{\rm eff}\simeq 1.1$ \citep{larson69, on98}. In MH1, MH5 and MH7, bumps in the density profile at a distance of a few tens of au from the centre indicate that secondary clumps have begun to form. The enclosed mass is less sensitive to fluctuations in the density and does not show any prominent features. The temperature remains at about $1000\,{\rm K}$ from about $1000$ to $10\,{\rm au}$, followed by an increase to between $1500$ and $2000\,{\rm K}$ on a scale of $1\,{\rm au}$. In MH6, HD cooling becomes important and prolongs the initial cooling phase to $n_{\rm H}\simeq 10^6\,{\rm cm}^{-3}$, after which the temperature increases more sharply compared to haloes in which HD cooling does not become important. This leads to a slower collapse of the cloud during which turbulent motions decay and a pronounced disc forms \citep{clark11a}. The molecular hydrogen fraction shows little scatter among the individual haloes and increases from $y_{\rm H_2}\simeq 10^{-3}$ at $\simeq 1000\,{\rm au}$ to fully molecular with $y_{\rm H_2}\simeq 0.5$ at a few tens of au. The radial velocity over the sound speed peaks on a scale of a few tens of au as the gas cools slightly and loses some pressure support.

\begin{figure*}
\begin{center}
\includegraphics[width=16cm]{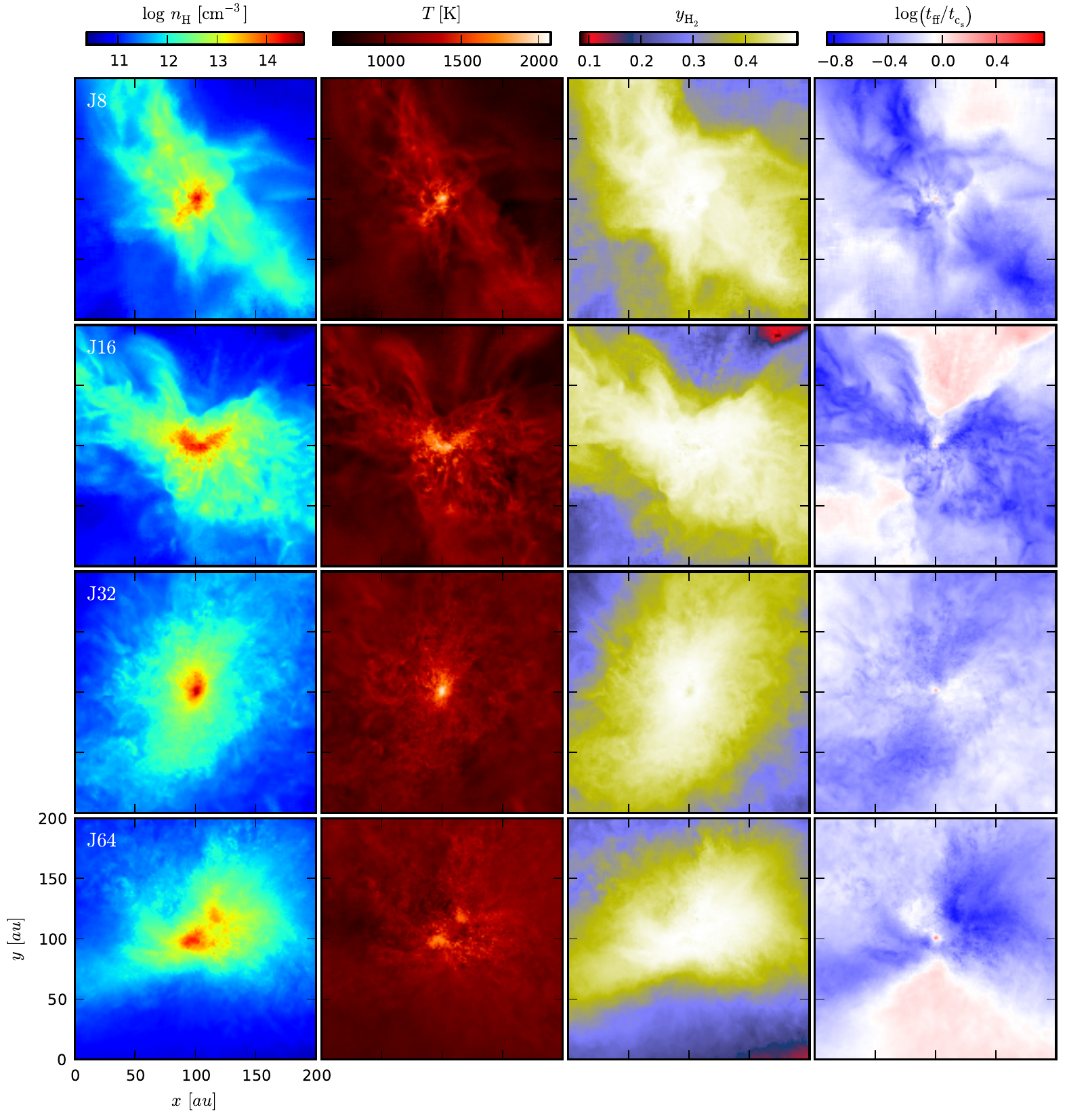}
\caption{From left to right: number density of hydrogen nuclei, temperature, H$_2$ fraction and free-fall time over sound-crossing time in cubes of side length $200\,{\rm au}$, centred on the densest cell in the halo and weighted with the mass and the square of the density of the cells traversed along the line of sight. From top to bottom: simulations employing $8$, $16$, $32$ and $64$ cells per Jeans length according to \citet{tna10} in MH1. Except for an increasing amount of small-scale structure, there is no apparent trend in any of the shown quantities as the resolution is increased.}
\label{fig_res_img}
\end{center}
\end{figure*}

\begin{figure*}
\begin{center}
\includegraphics[width=16cm]{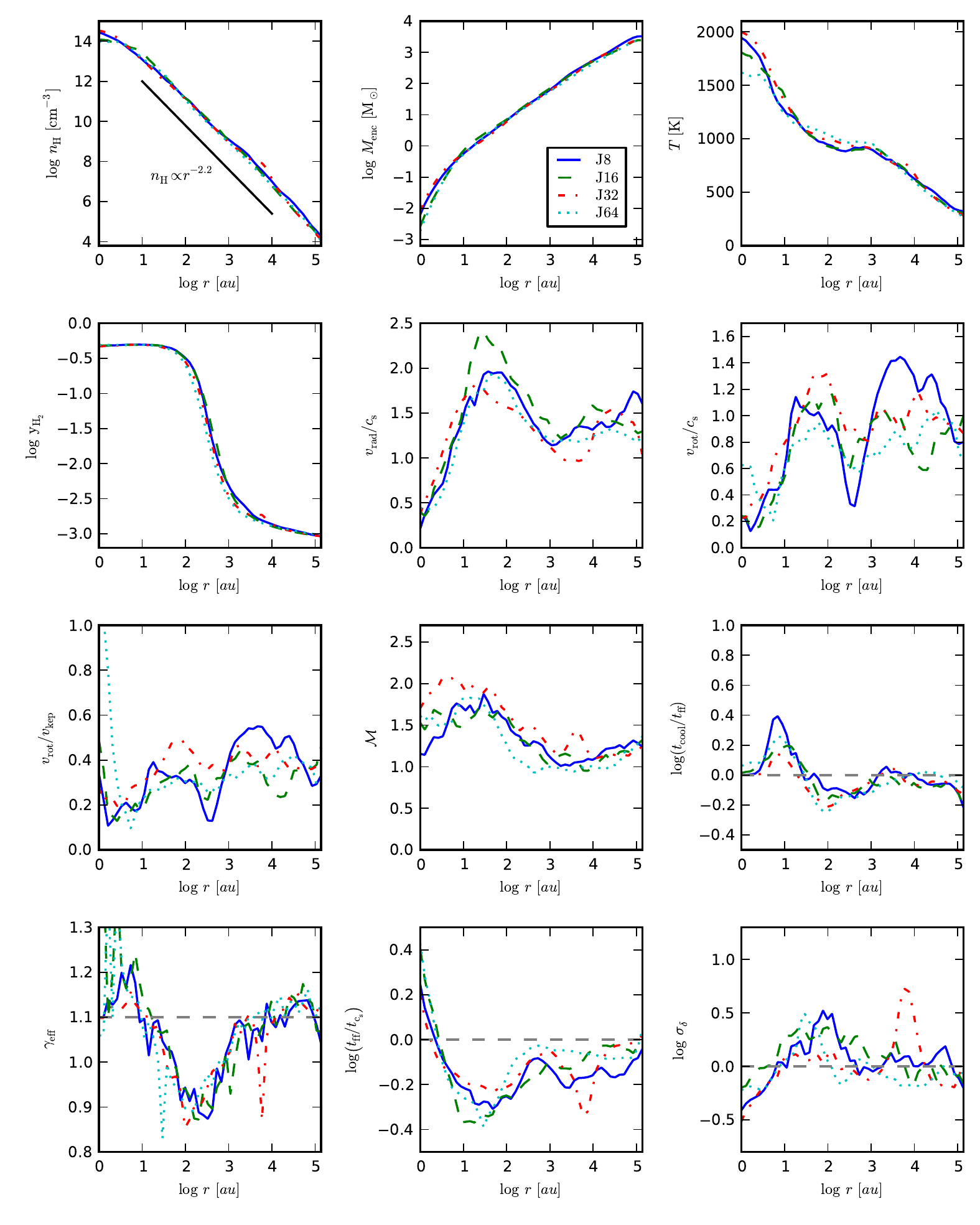}
\caption{From top left to bottom right: number density of hydrogen nuclei, enclosed gas mass, temperature, H$_2$ fraction, radial velocity over sound speed, rotation velocity over sound speed, rotation velocity over Keplerian velocity, turbulent Mach number, cooling time over free-fall time, effective equation of state, free-fall time over sound-crossing time and rms density contrast versus radius for $8$, $16$, $32$ and $64$ cells per Jeans length according to \citet{tna10} in MH1. The solid black line in the top left panel denotes an $r^{-2.2}$ density profile. The grey dashed lines denote a cooling time equal to the free-fall time, an effective equation of state of $1.1$, a free-fall time equal to the sound-crossing time and an rms density contrast of unity. Qualitatively, the profiles are similar among the various simulations, but also display a substantial amount of scatter. We find no clear trend in any of the profiles as the resolution is increased.}
\label{fig_res_radial}
\end{center}
\end{figure*}

All clouds are substantially rotationally supported. The average rotation velocity is of the order of the sound speed and comparable to the radial velocity. The rotation velocity over the Keplerian velocity fluctuates by about a factor of $2$ in the course of the collapse of a cloud, and has a mean value of $\simeq 1/3$ among the haloes. In MH6, the cooling time briefly increases to nearly an order of magnitude above the free-fall time, and the prominent disc visible in Fig.~\ref{fig_multi_dens} forms. This is reflected by an increase of the rotation velocity over the Keplerian velocity to $\simeq 0.6$ on a scale of $\simeq 1000\,{\rm au}$. The turbulent Mach number is generally close to unity, implying that the clouds are in nearly equal parts supported by thermal pressure, rotation and turbulence. Below $\simeq 100\,{\rm au}$, the turbulence transitions into the supersonic regime with $\mathcal{M}\simeq 1.5$. The turbulent support of the cloud is also apparent from the filamentary appearance of the gas shown in Fig.~\ref{fig_multi_dens}.

Following the rapid increase of the H$_2$ fraction on a scale of $\simeq 1000\,{\rm au}$, the cooling rate increases and the cooling time drops below the free-fall time (with the exception of MH6). The effective equation of state of the gas decreases from $\gamma_{\rm eff}\simeq 1.1$ to $\simeq 0.9$ on a scale of $\simeq 100\,{\rm au}$, and the free-fall time drops below the sound-crossing time, with MH1, MH5 and MH7 showing the most pronounced dip. During this phase, the gas becomes gravitationally unstable and the rms density contrast increases to well above unity. In MH5 and MH7, the density contrast increases by approximately an order of magnitude. The formation of highly gravitationally unstable regions in MH1, MH5 and MH7 is evident from Fig.~\ref{fig_multi_collapse}, which shows the free-fall time over the sound-crossing time in the central $200\,{\rm au}$ of the box.

In the following, we investigate whether the secondary clumps visible in Figs.~\ref{fig_multi_dens} and \ref{fig_multi_collapse} have accumulated enough mass to collapse in their own right, and whether they will collapse before they are accreted by the primary clump. The top panel in Fig.~\ref{fig_mbe} shows the enclosed mass over the locally estimated Bonnor--Ebert mass \citep{ebert55, bonnor56}, centred on the minimum in the free-fall time over the sound-crossing time as identified in Fig.~\ref{fig_multi_collapse}. The Bonnor--Ebert mass is computed as the mass-weighted average among the cells within a given radius according to
\begin{equation}
M_{\rm BE}\simeq 15\,{\rm M}_\odot\,\left(\frac{n_{\rm H}}{{\rm cm}^{-3}}\right)^{-1/2}\left(\frac{T}{{\rm K}}\right)^{3/2}\mu^{-3/2}\gamma^2.
\end{equation}
The bottom panel shows the free-fall time over the accretion time. The free-fall time is computed by using the mass enclosed within a given radius, and the accretion time by $t_{\rm acc}=d/v_{\rm rad}$, where $d$ denotes the distance of the clump from the centre of the cloud (vertical lines in Fig.~\ref{fig_mbe}) and $v_{\rm rad}$ the net radial velocity of the mass enclosed within a given radius with respect to the centre of the cloud.

From this figure, it is evident that only the clouds in MH1, MH5 and MH7 are likely to fragment during the initial free-fall phase. The enclosed mass exceeds the Bonnor--Ebert mass at small enough radii that the clumps are Jeans unstable in their own right, which is not the case in the other haloes, where there is no clear distinction between the primary and secondary clumps. Furthermore, since the free-fall time falls below the accretion time at radii significantly smaller than the separation of the clumps from the centre of the cloud, the clouds in MH1, MH5 and MH7 will likely collapse before they are accreted. However, the separation of the clumps is only of the order of a few tens of au -- in contrast to the $800\,{\rm au}$ that were found in \citet{tao09}. Based on the results of \citet{greif12}, it thus appears likely that the secondary clumps will merge well before radiation feedback can evacuate enough mass to alleviate the strong gravitational torques that are present in the disc. On the other hand, the secondary clump found in \citet{turk12} has a much higher chance of surviving and forming a separate Population~III star.

\begin{figure}
\begin{center}
\includegraphics[width=8cm]{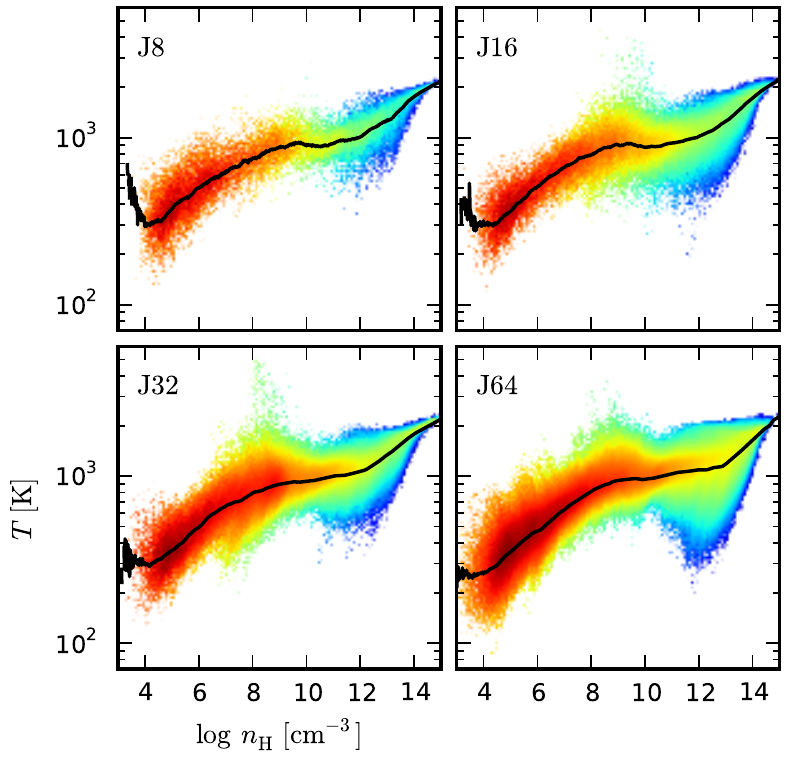}
\caption{Temperature versus number density of hydrogen nuclei for $8$, $16$, $32$, and $64$ cells per Jeans length according to \citet{tna10} in MH1. The logarithm of the mass per bin over the total mass in the box is colour coded from blue (lowest) to red (highest). The solid lines show the mass-weighted average value. As in Figs.~\ref{fig_res_img} and \ref{fig_res_radial}, we find no clear trend as the resolution is increased.}
\label{fig_res_pspace}
\end{center}
\end{figure}

\subsection{Resolution study}

Previous studies of Population~III star formation have argued that the nature of primordial gas clouds depends sensitively on the employed resolution \citep[e.g.][]{turk12}. We investigate whether the resolution requirements found in these simulations also apply to the simulations presented here. To this end, we employ between $8$ and $64$ cells per Jeans length according to \citet{tna10}, using the halo that was described in detail in Section~3.1 (MH1).

In Fig.~\ref{fig_res_img}, we show the number density of hydrogen nuclei, temperature, H$_2$ fraction and free-fall time over sound-crossing time in the central $200\,{\rm au}$ of the box. The morphology of the gas is highly irregular in all simulations, and the amount of small-scale structure  increases as the resolution is increased. However, we do not find any other convincing trend in the density, temperature, H$_2$ fraction and free-fall time over sound-crossing time. Gravitationally unstable regions are visible in all simulations, but their size or abundance does not systematically increase or decrease as the resolution is increased.

In Fig.~\ref{fig_res_radial}, we show radial profiles of the physical quantities discussed in Section~3.2. Qualitatively, the profiles agree well with each other, but there is also a substantial amount of scatter. As in Fig.~\ref{fig_res_img}, we find no clear trend in any of the shown physical quantities as the resolution is increased. Finally, in Fig.~\ref{fig_res_pspace}, we show the distribution of the gas in density and temperature, but again there is no clear trend as the resolution is increased. It appears that the random nature of self-gravitating collapse and the non-linear rate equations that govern the chemical and thermal evolution of the cloud dominate over all other trends that may be present. In particular, the differences between simulations with different resolution are similar to the differences between simulations of different haloes.

\section{Summary and conclusions}

We have investigated the chemothermal and gravitational stability of primordial, star-forming clouds with a suite of nine high-resolution simulations. At densities $n_{\rm H}\ga 10^8\,{\rm cm}^{-3}$, three-body reactions rapidly increase the H$_2$ fraction and trigger potential runaway cooling, which is counteracted by the chemical heating of the gas due to the release of the binding energy of H$_2$, and the increasing optical depth to H$_2$ line emission. The competition between these processes creates a characteristic dip of the cooling time over the free-fall time on a scale of $\simeq 100\,{\rm au}$, and in consequence the free-fall time falls below the sound-crossing time on a scale of a few tens of au. As a result, gravitational instability and the collapse of secondary clumps may be triggered. In three of the nine haloes investigated, enough mass accumulates in the secondary clumps that they will likely collapse before they are accreted by the primary clump.

In addition to the main simulation suite, we have performed a resolution study to investigate how our results depend on the number of cells employed per Jeans length. We find no convincing trend in the appearance of the cloud or other physical quantities as the resolution is increased. Instead, the turbulent nature of self-gravitating collapse and the non-linear rate equations that govern the chemical and thermal evolution of the cloud appear to dominate over all other trends. The differences between simulations with different resolution are comparable to the differences between simulations of different haloes.

Our results differ from previous studies in a few aspects. In the simulation of \citet{yoshida06b}, the primordial gas cloud did not fragment, while we here find that fragmentation may occur. This could simply be a consequence of low-number statistics, since \citet{yoshida06b} investigated a single halo. However, it may also be related to systematic differences in the morphologies of the clouds investigated. In the present study, the clouds show a more filamentary appearance due to transonic turbulence, which is in better agreement with the results of \citet{tao09}. However, in terms of fragmentation, \citet{tao09} found that only one out of five haloes fragmented. The scale of the fragmentation was $\simeq 1000\,{\rm au}$ instead of a few tens of au, and was attributed to the rotation of the cloud instead of the chemothermal instability. A possible cause for this difference may be the density-dependent fitting function \citet{tao09} used for the calculation of the photon escape fraction, which has been shown to yield less efficient cooling than the Sobolev method \citep{turk11}. A final difference is that the gas clouds in \citet{turk12} become more rotationally supported when less than $32$ cells per Jeans length are used. We do not find such a trend even when the resolution is decreased to only eight cells per Jeans length, which is close to the bare minimum required to resolve self-gravitating collapse \citep{truelove98, greif11a}.

We have made a number of simplifying assumptions concerning the setup of the simulations. In particular, we have not taken supersonic streaming motions into account, which have a substantial effect on the mass scale and virialization of the haloes in which Population~III star formation occurs \citep{th10, greif11b, mkc11, sbl11b, nyg12, nyg13, om12}. We further neglect the potential heating of the gas by self-annihilating DM \citep{freese08, iocco08, ripamonti10, smith12}. Instead of performing full radiation transfer for H$_2$ line emission, we resort to approximate, on-the-spot methods that likely reduce the cooling efficiency of the gas. Furthermore, some of the employed chemical rate equations are highly uncertain \citep{ga08, turk11}. Finally, we neglect the influence of magnetic fields, which might become dynamically important during the initial collapse \citep{machida06, mmi08, xu08, schleicher09, schleicher10, sur10, federrath11, schober12, turk12}.

In concluding, we note that while not all star-forming clouds in minihaloes are expected to fragment during their initial collapse, this by no means precludes fragmentation at a later stage, when the gas becomes rotationally supported in a Keplerian disc, and the infall rate is significantly reduced compared to simple one-dimensional estimates \citep{greif12}. The chemothermal instability and the resulting gravitational instability may therefore have substantially more time to develop than is apparent from the limited period of time simulated here.

\section*{Acknowledgements}
THG would like to thank Tom Abel and Naoki Yoshida for stimulating discussions that helped improve this work. The simulations were carried out at the Rechenzentrum Garching (RZG) and the Texas Advanced Computing Center (TACC) under XSEDE allocation AST130020 (for THG) and project A-astro (for VB). VB acknowledges support from NSF through grant AST-1009928.


\end{document}